\documentclass[reprint,floatfix,amsmath,amssymb,aps,showpacs,prl,superscriptaddress]{revtex4-1}
\usepackage[]{graphicx}
\usepackage[caption=false]{subfig}
\usepackage[utf8]{inputenc}
\usepackage[ngerman,english]{babel}
\usepackage{soul}\setstcolor{magenta}
\usepackage[hidelinks,pdfview={FitH},pdfstartview={FitH},bookmarksopen={false}]{hyperref}
\usepackage{color}
\usepackage{chapterbib}
\begin{document}
%\date{\today}

\title{Nucleation in sheared granular matter}

\author{Frank Rietz}
\email[email: ]{frank.rietz@gmx.net}
\affiliation{Max-Planck-Institute for Dynamics and Self-Organization G\"ottingen, 37077 G\"ottingen, Germany}
\affiliation{Institute for Multiscale Simulation, Friedrich-Alexander-Universit\"at Erlangen-N\"urnberg (FAU), 91052 Erlangen, Germany}
\affiliation{Department of Nonlinear Phenomena, University Magdeburg, Universit\"atsplatz 2, 39106 Magdeburg, Germany}
\affiliation{Department of Pattern Formation, University Magdeburg, Universit\"atsplatz 2, 39106 Magdeburg, Germany}
\author{Charles Radin}
\affiliation{Department of Mathematics, University of Texas at Austin, Austin, TX, USA}
\author{Harry L. Swinney}
\affiliation{Center for Nonlinear Dynamics and Department of Physics, University of Texas at Austin, Austin, TX, USA}
\author{Matthias Schr\"oter}
\email[email: ]{matthias.schroeter@ds.mpg.de}
\affiliation{Max-Planck-Institute for Dynamics and Self-Organization G\"ottingen, 37077 G\"ottingen, Germany}
\affiliation{Institute for Multiscale Simulation, Friedrich-Alexander-Universit\"at Erlangen-N\"urnberg (FAU), 91052 Erlangen, Germany}

\begin{abstract}
We present an experiment on crystallization of packings of macroscopic granular spheres. This system is often considered to be a model for thermally driven atomic or colloidal systems. Cyclically shearing a packing of frictional spheres, we observe a first order phase transition from a disordered to an ordered state. The ordered state consists of crystallites of mixed FCC and HCP symmetry that coexist with the amorphous bulk. The transition, initiated by homogeneous nucleation, overcomes a barrier at 64.5\% volume fraction. Nucleation consists predominantly of the dissolving of small nuclei and the growth of nuclei that have reached a critical size of about ten spheres. 
\end{abstract}
\maketitle
Packings of spheres show interesting features such as phase transitions between disordered and ordered states, and can be useful in the study of amorphous atomic configurations \cite{Finney2013}. Examples include thermal colloidal packings \cite{Pusey1986,Gasser2001,Liber2013},
packings of macroscopic granular spheres 
\cite{Scott1960_Scott1969_Finney1970,Finney2013,Bernal1964_Nicolas2000_Mueggenburg2005,Panaitescu2012,Francois2013,Hanifpour2014_Hanifpour2015_Saadatfar2017,Tsai2003_Daniels2006,Chen2006_Slotterback2008,Knight1995_Richard2005,Schroeter2005,Berryman1983,Komatsu2015}, and simulations of the mathematical hard sphere model  \cite{Royer2015,Kapfer2012,Mickel2013,Anikeenko2007_Baranau2014,Jin2010,Wood1957_Alder1957_Hoover1968,Auer2004,Sanz2011_Valeriani2012,Yanagishima2017,Sanz2014}.
The behavior of such systems is determined by the fraction of space filled by the spheres, their packing fraction $\phi$.

The hard sphere model exhibits an entropically driven first order phase transition. Disordered fluid states are observed below the freezing density of $\phi=0.495$ and crystalline ordered states appear above the melting density of $\phi=0.545$, with coexistence of the two phases for intermediate densities \cite{Wood1957_Alder1957_Hoover1968}.

Granular spheres can also be packed in disordered and ordered states. In contrast 
to colloidal packings and the mathematical hard sphere model, granular packings are characterized by the existence of permanent contacts between the particles. Most experimental protocols for increasing the bulk volume fraction, such as vertical shaking \cite{Knight1995_Richard2005}, centrifugation \cite{Liber2013}, thermal cycling \cite{Chen2006_Slotterback2008} and sedimentation \cite{Schroeter2005}, do not achieve an ordered state from an initial disordered state.
The ``random close packed state'' (RCP)  is  used operationally to describe the highest density state achieved by these methods. The RCP volume fraction is in the range $0.635 <\phi< 0.655$ \cite{Scott1960_Scott1969_Finney1970,Berryman1983,Anikeenko2007_Baranau2014,Kapfer2012}, 
about $15\%$ lower than the densest possible packing of ordered face centered cubic (FCC) or hexagonal close packing (HCP), which each have a volume fraction $\phi = \pi/\sqrt{18} \approx 0.74$ \cite{Hales2012}. 

Ordered clusters of granular spheres have been obtained with a system density $\phi_{global}$ in the range 0.64--0.74 by multidimensional shaking \cite{Francois2013,Hanifpour2014_Hanifpour2015_Saadatfar2017}, cyclic shear \cite{Bernal1964_Nicolas2000_Mueggenburg2005,Panaitescu2012}, and shear in a Couette cell \cite{Tsai2003_Daniels2006}.
By analogy with the freezing-melting transition in the hard sphere model, the emergence of growing crystallites found in our granular experiment  can be interpreted in terms of nucleation and a first order phase transition \cite{Radin2008_Aristoff2010,Jin2010}. 

In our experiment we compact a granular packing by shearing.   With increasing shear cycles, a well-defined plateau emerges at phase transition density $\phi_{global}=0.645$, which is in the range of densities associated with RCP \cite{Scott1960_Scott1969_Finney1970,Berryman1983,Anikeenko2007_Baranau2014,Kapfer2012}. Such a plateau was not reported in a previous experiment that used a setup similar to ours, and the growth rate of nuclei that we find  differs qualitatively from that found in the previous experiment \cite{Panaitescu2012}. 

\begin{figure}[b!]
\includegraphics[width=\columnwidth]{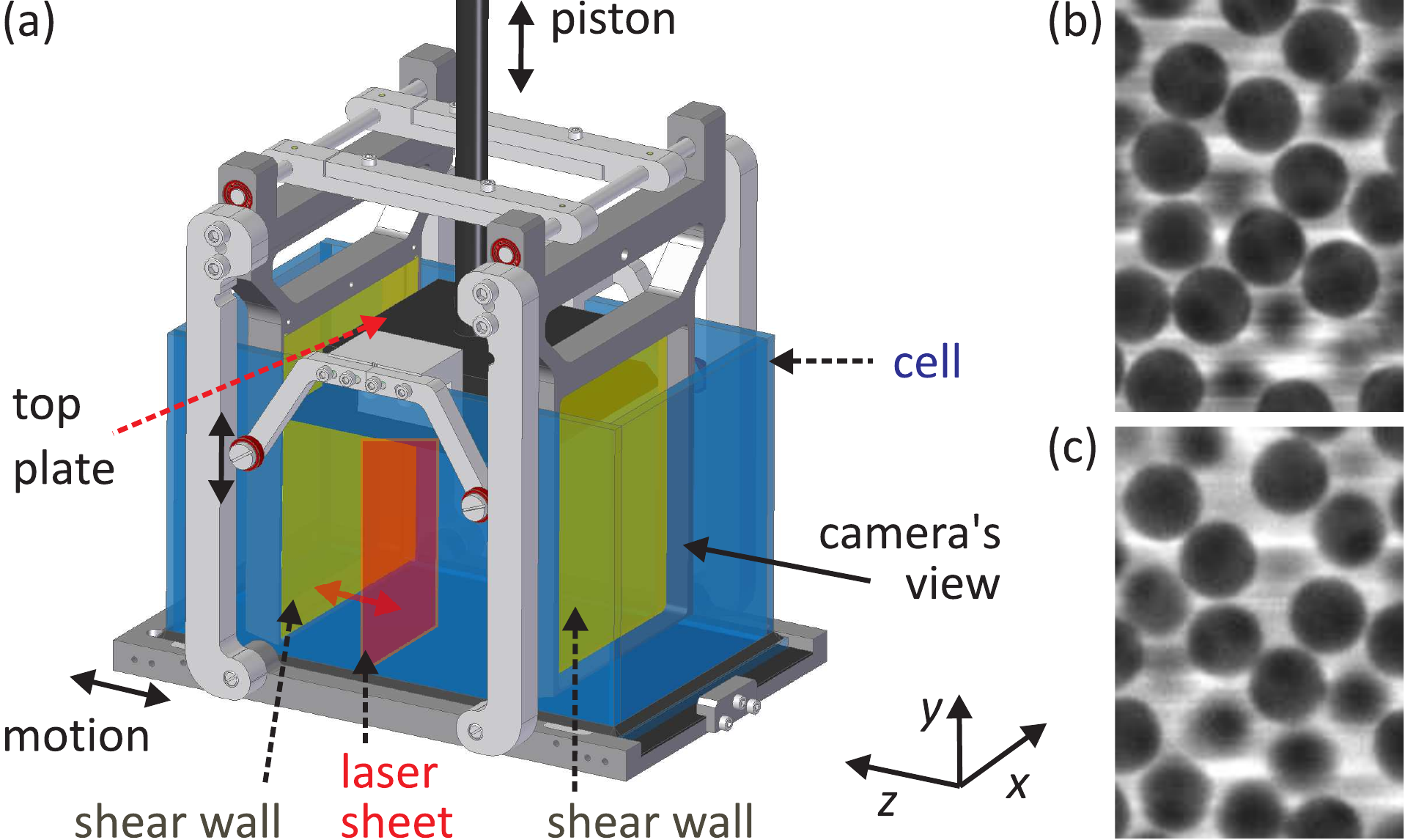}
\caption{\label{fig2}
Precision spheres are contained in a cubical volume with opposite side walls (lime green) that produce oscillating shear. The cell is filled with a liquid that is index matched to the beads. The top plate is levered to the side walls and is mounted on a piston that is constrained to move vertically. Shear is produced by periodically oscillating the cell bottom in the $z$ direction with a period of 2\,s. A laser sheet illuminates a slice in the $x-y$ plane, as illustrated by contrast enhanced images in (b) and (c); (b) shows a layer near the camera, and (c) shows a layer much further from the camera. Note that the image quality in both layers is comparable.}
\end{figure}

Our experiment uses a cubical shear cell with side length 10.5 cm (Fig.~\ref{fig2}(a)). Nucleation from the side walls is suppressed using half-spheres glued to the walls at random positions (glue: \foreignlanguage{ngerman}{Vitralit 7562, Panacol}).  The cell is filled with 49,400 precision BK7 glass spheres of diameter 3\,$\pm0.0025$\,mm and density 2.51\,g/cm$^3$ (size tolerance given by manufacturer, \foreignlanguage{ngerman}{Worf Glaskugeln GmbH}). The top plate applies a pressure of 2.1\,kPa on the bed, and the pressure increases downward to 3.1\,kPa at the bottom.  The cell is sheared by sinusoidally tilting opposite vertical shear walls by $\pm0.6^{\circ}$ about axes indicated by red circles about half-way up opposite side walls (Fig.~\ref{fig2}(a)); the total peak-to-peak oscillation displacement at  the bottom of the cell is about one-third of a sphere diameter. 

The glass spheres are index matched \cite{Dijksman2012} to a mixture of phthalate esters (Cargille Laboratories, identifier code 1160, $\rho=1.1$\,g/cm$^3$, $\nu=41$\,cSt) with a dissolved fluorescent dye (oxazine 750 perchlorate, $c=10$\,mg/l). Hydrodynamic interactions are negligible for our small amplitude shear with a period of 2\,s, and sphere deformation is also negligible. The spheres, half-spheres, and shear cell walls are made of BK7 optical glass and are index matched with the liquid mixture in a temperature controlled environment ($n=1.5198 \pm 0.0001 $ at $\lambda=589$\,nm and $T=22.9\pm0.1^\circ$C).

The  shearing process is periodically stopped to measure the packing using an horizontally translated vertical laser light sheet ($\lambda$=658\,nm, $P$=75\,mW). Fluorescent light is imaged simultaneously by the camera (Figs.~\ref{fig2}(b) and 1(c)).
The image slices are combined to form a three-dimensional volume. The measurements presented here are from 618 scans made during a run with $1.955\times 10^6$ cycles.  In the course of the two month long run 10\% of the spheres escaped from the sample cell through a gap between one of the shear walls and a side wall, and this possibly increased the mobility of spheres farther inside the packing, thereby enhancing their crystallization.

Positions of the 20,000 spheres at least 3 diameters from any wall are detected by convolution with a template. Then the peak of the correlation map in the ($x$,$y$,$z$) directions is determined with (37,37,30) pixel/diameter resolution using a 3-point Gauss estimator \cite{Raffel2007}.
This yields the position of each particle. From the pair correlation function we can determine, for more than 99\% of the particles, the position of each particle to less than 2\% of a sphere diameter (see Supplemental Material \cite{supp}). 

\begin{figure}[b!]
\includegraphics[trim=19 0 0 0,clip,width=\columnwidth]{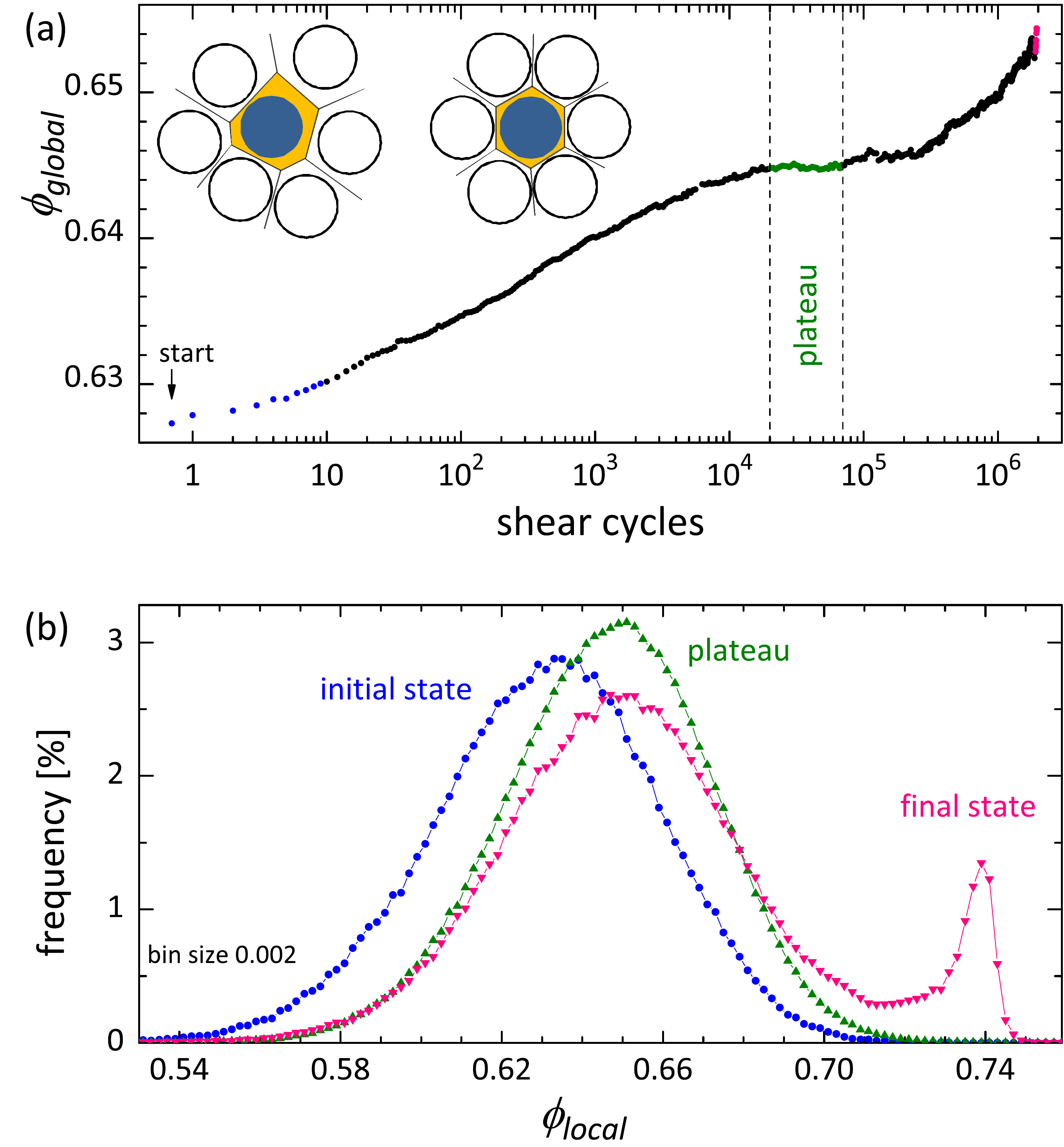}\caption{\label{fig1}
The formation of the crystalline phase shows features of a first order phase transition.
(a)~Starting from a loose packing, the global volume fraction $\phi_{global}$ increases logarithmically with the number of shear cycles until it reaches a plateau at the density $\phi_{global}=0.645$. After approximately 50,000 more shear cycles, $\phi_{global}$ increases again due to the formation of  crystalline regions inside the sample. The 2D diagrams on the upper left show spheres (blue) and their Voronoi cells (yellow) for cases with Voronoi neighbors that are disordered loosely packed (left) and symmetric densely packed (right). (b)~Histograms of local volume fractions reveal the coexistence of crystalline and amorphous regions inside the sample. Below the transition there is only a single peaked distribution, which shifts towards higher densities until the plateau is reached. After the onset of crystallization the previous peak population diminishes and a new peak appears at $\phi_{local}=0.74$. The histogram for the plateau is the average of 32 consecutive scans; elsewhere the histogram is the average of 10 scans. Colors in (b) correspond to the colored data points in (a). 
}\end{figure}

For each sphere there is a Voronoi cell consisting of all points closer to that sphere than to any other sphere in the sample (see insets in Fig.~\ref{fig1}(a)). $\phi_{local}$ is then the ratio of the sphere volume to the volume of its Voronoi cell. The mean volume fraction of the whole sample, $\phi_{global}$, is given by the harmonic mean of all $\phi_{local}$ values \cite{Weis2017}.
 
The angular order between spheres sharing a face of their Voronoi cells is characterized by a weighted version of the order parameter $q_6$ \cite{Mickel2013}.
A sphere is called crystalline if it is densely packed and its neighbors are ordered, i.e., $\phi_{local}>0.72$ {\it and} $q_6$ is either in the range $q_6$(FCC)=0.575$\pm$0.020 or $q_6$(HCP)=0.485$\pm$0.020. Other spheres are called amorphous. A nucleus is a connected set of crystalline spheres, each sharing at least one Voronoi face with another sphere in the set. Some authors suggest different definitions for local order \cite{Kapfer2012,Faken1994_Bargiel2001_Stukowski2012_Tanaka2012_Leocmach2013}, but our results do not depend qualitatively on the choice of definition or threshold  (see Supplemental Material \cite{supp} and a recent review \cite{Reinhart2017}).  The distance of a nucleus to another single sphere or nucleus is given by the shortest distance between the sphere centers. 

\begin{figure}[b!]
\includegraphics[width=\columnwidth]{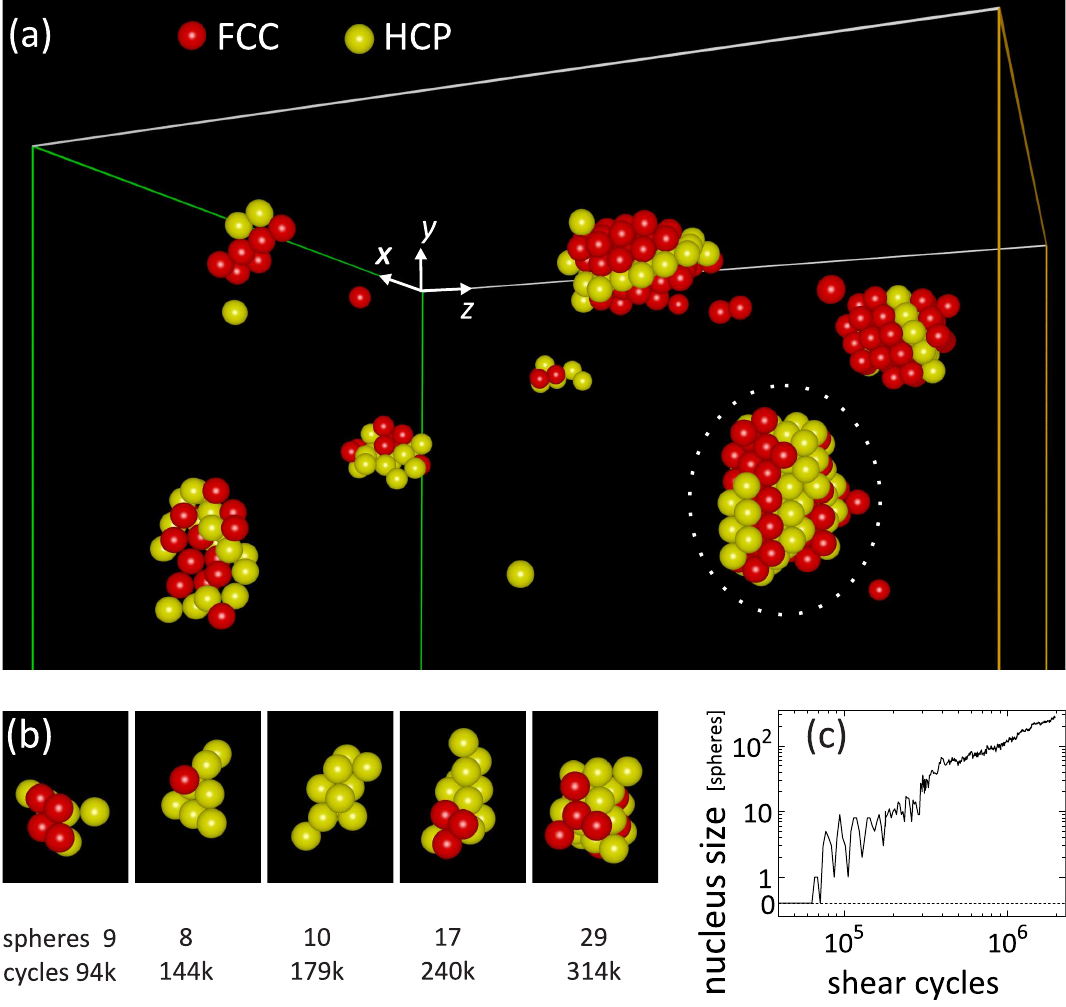}\caption{\label{fig5}	
(a)~This cross section of the cell after $10^6$ cycles shows crystalline regions in the cell interior; the amorphous phase is not shown. The colors of the nucleated spheres indicate crystal type. The wire frame indicates the inner part of the shear cell that is used for evaluation.
(b)~The nucleus that started to grow first (dotted ellipse in (a)) fluctuated in shape and size until a stable seed was reached at about $3 \times 10^5$ cycles. (c)~Time evolution of the nucleus in (b). A movie in the Supplemental Material \cite{supp} illustrates the nucleation in the shear cell.}
\end{figure}

During densification we observe three distinct regions, as can be seen in Fig.~\ref{fig1}(a). The packing starts from a disordered state and compacts approximately logarithmically with time for about 20,000 cycles. The compaction then slows to a stop, and the second region, a plateau  ($\phi_{global}=0.645$), emerges and persists for about 50,000 cycles. Then the first growing nucleus appears, indicating a first order phase transition to the third distinct region. The volume fraction slowly begins to increase as shearing continues, and nuclei increase in number and size but have no preferred orientation of their hexagonal layers; an intermediate state of the system is shown in Fig.~\ref{fig5}(a). The first growing nucleus is shown in Fig.~\ref{fig5}(b), which illustrates that nuclei fluctuate in shape, size, and crystal symmetry as they grow, as can be seen in the movie in the Supplemental Material \cite{supp}.

All nuclei start their growth at least 10 sphere diameters distance away from any wall. By the end of the experiment, after two million shear cycles, nuclei with up to $\sim 600$ spheres are present, and  9\% of all spheres in the analyzed volume are in crystallites, which have FCC or HCP symmetry with approximately equal probability; no icosahedral symmetry was observed (see Supplemental Material \cite{supp}).

Histograms of {\it local} densities for amorphous packings have a single peak and are approximately symmetrical about that peak, as Fig.~\ref{fig1}(b) illustrates. During compaction the peak narrows slightly and shifts to higher densities.  Beyond the first order phase transition a second peak emerges at $\phi_{local}=0.74$, the density of densest packed arrangements.

\begin{figure}[b!]
\includegraphics[trim=2 0 3 0,clip,width=\columnwidth]{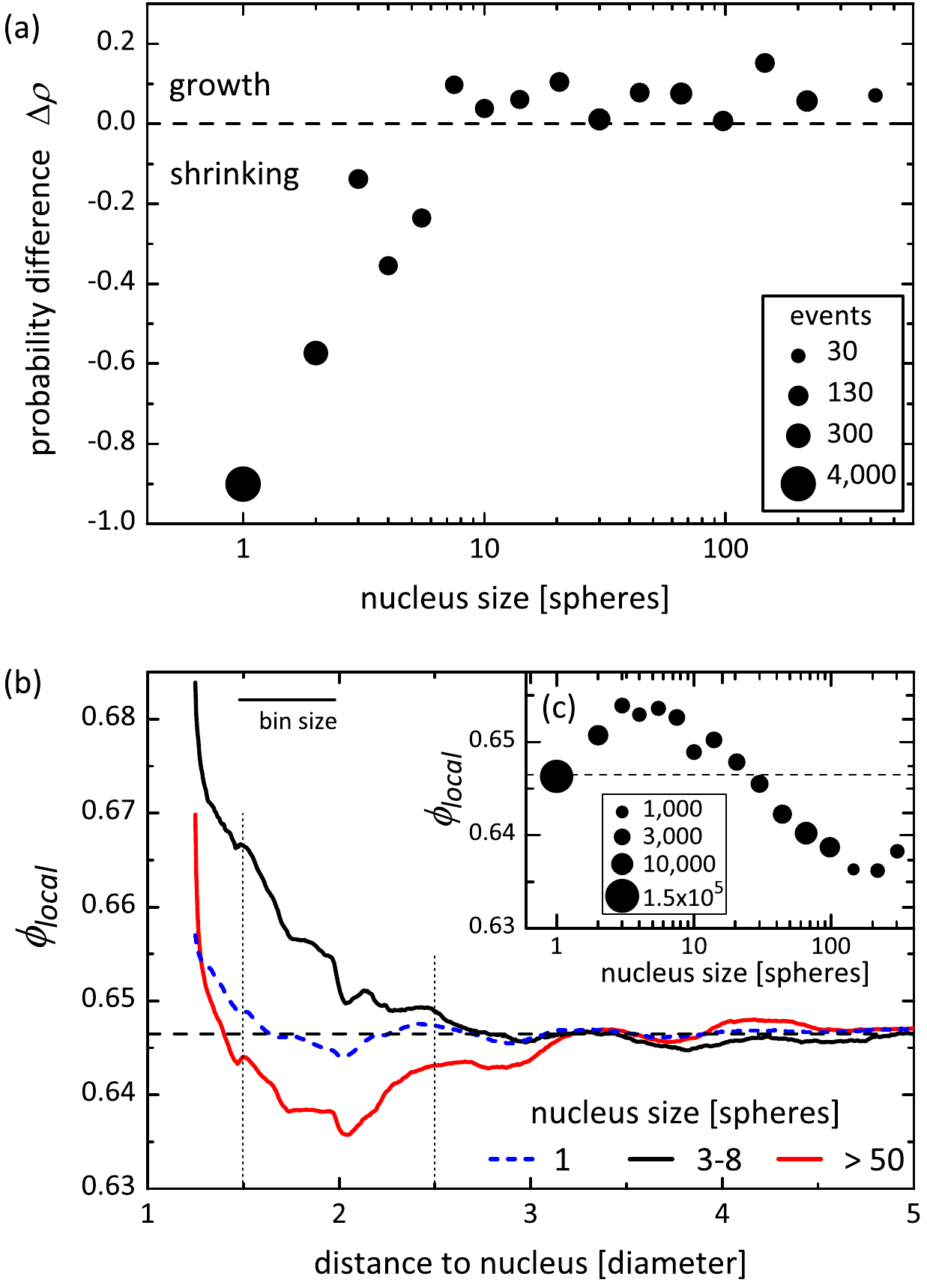}\caption{\label{fig4}	
(a)~The growth (shrinking) probability of a nucleus depends on its size: nuclei with about ten or more spheres predominately grow ($\Delta\rho>0$).  (b)~Local volume fraction as a function of distance from the surface of a nucleus. Large nuclei (red) show a minimum at a distance of about 2 sphere diameters, but there is no comparable minimum for subcritical nuclei (black), and very small nuclei have a minor effect on the surroundings (dashed). (c)~The packing density near the surface of a nucleus -- in the range in (b) between the vertical dotted lines -- is higher for the small nuclei that tend to dissolve than for the large nuclei that tend to grow (see Figures S3-S12 in the Supplementary Material \cite{supp}). The horizontal dashed line in (b) and (c) indicates the amorphous plateau far from any nucleus. 
}\end{figure}

The end of the densification plateau is identified by the emergence of a nucleus in the interior region; subsequently all nuclei are tracked for each successive scan. Nuclei with fewer than about ten spheres are found to shrink more often than grow, while growth gradually becomes more probable for nuclei with more than ten spheres, that is, $\Delta\rho=(\rho_{grow}-\rho_{shrink}) > 0$, where $\rho$ is the probability to grow or shrink, respectively \cite{Gasser2001,Panaitescu2012} (Fig.~\ref{fig4}(a)). The critical size at which the difference of the  probabilities $\Delta\rho$ becomes positive does not depend on the definition of local crystallinity (see Supplemental Material \cite{supp}). 

% \F{In an experiment with thermal colloids}\CR{, a very different system,} \F{the size of critical nuclei has been measured to be 60--160 spheres \cite{Gasser2001}.}

Our main result is the observation of nucleation centers throughout the interior of a granular packing.  The emergence of a plateau at a well-defined packing fraction, $\phi_{global}=0.645$, indicates the onset of a first order phase transition.  A previous experiment used a setup similar to ours and observed heterogenous nucleation at the cell walls but no nucleation in the interior of the cell \cite{Panaitescu2012}.  Another experiment used shaking in a spherical container which may or may not have suppressed heterogeneous nucleation; however, the spatiotemporal evolution of the nuclei was not determined in this study (see \cite{Francois2013,Hanifpour2014_Hanifpour2015_Saadatfar2017} and Supplementary Material in \cite{Francois2013}).

Our experiment analyzes static granular matter between episodes of cyclic shearing, which increase the density.  Our data are snapshots, measurements of static packings. When the snapshots are viewed consecutively, as in a movie, nuclei emerge, grow, and shrink until after sufficient shearing cycles, nuclei of critical size are created that then grow indefinitely under further shearing (see movie in Supplementary Material \cite{supp}). A nucleation theory should explain why shearing at low amplitude and low frequency, in the presence of pressure and gravity, leads to such a sequence of static granular configurations. This is distinct from the usual nucleation theory, which shows that a thermal or Brownian dynamics, starting from a supercooled state, leads to the creation, growth, and shrinking of nuclei.

The observed compaction of a granular bed under gravity and pressure can be described by a simple mechanical picture.  A bed of frictional hard spheres increases in density if subjected to repeated small disturbances that are introduced, for example, by shearing, shaking, or fluidization.  The repeated shear cycles in our experiment break some of the force chains that form a skeleton supporting the bed, and this leads to compaction \cite{Yanagishima2017}. However, this simple picture does not address the homogeneous nucleation seen in our experiment (Fig.~\ref{fig4}(a)); very dense crystallite clusters ($\phi=0.74$) form, grow, and dissolve without the help of a flat wall.  Figure~\ref{fig4}(b) shows that near a small nucleus the volume fraction is slightly higher than the background volume fraction, which suggests from a mechanical argument  that under a confining pressure small nuclei should grow. However, Fig.~\ref{fig4}(a) reveals that this does not happen. Any theory for nucleation in our system would have to account for these results in Fig.~\ref{fig4}.

% Our Fig. 4(b) shows that these isolated dense clusters are in a bath of grains at the random closed packed volume fraction, and that small clusters -- with about ten or fewer particles -- tend to dissolve, while larger clusters tend to grow. These results should provide a guide for the development of a theory that would explain the observed homogeneous nucleation.

Our results are qualitatively different from the  crystallization dynamics observed in constant volume molecular dynamics simulations of supercooled hard spheres: the existence of a nucleation barrier in our system disagrees with the autocatalytic growth of crystalline regions observed in \cite{Sanz2011_Valeriani2012}. Further, we do not observe burst-like growth events as seen in mature glasses \cite{Yanagishima2017,Sanz2014}.  These differences indicate that nucleation in our sheared granular system is governed by its own dynamics.

In conclusion, we have found that small amplitude cyclic shear of a bed of spherical particles under gravity and pressure leads to compaction until a well-defined random close packed volume fraction is reached. Then after many more cycles (50,000 in our experiment), clusters with HCP and FCC symmetry emerge, and these crystallites grow if they contain about ten or more particles.  Previous experiments  \cite{Bernal1964_Nicolas2000_Mueggenburg2005,Tsai2003_Daniels2006, Panaitescu2012,Francois2013,Hanifpour2014_Hanifpour2015_Saadatfar2017} had observed nucleation centers on flat confining walls, but our experiment shows  nucleation throughout the interior of the cell.  Further, the present experiment on 49,400 spheres indicates that future numerical simulations of sheared packings should extend well beyond a recent study of 2,000 frictional grains cyclically sheared for 2,000 cycles \cite{Royer2015}.\\

% \MS{An interesting aspect of our result emerges when considering that our sample is under a confining pressure. Figure ~\ref{fig4}(b,c) shows that the density at the interface of small nuclei is larger than the bulk density. In consequence, it should be energetically favorable for the system to create more interface, which means that small nuclei would grow. Given that this is not the case (Fig.~\ref{fig4}(a)) we speculate that another mechanism, such as geometric inhibition \F{(arrest?)}, might prohibit the growth of small nuclei.} \MS{\it Arrest implies fixation, inhibition implies avoidance. The latter is imho the better picture.}

% \F{\it Solely considering Fig.4(b,c) means 2 things: (I) small nuclei should be initially motivated to grow, (II) nuclei should stop to grow when they reach a size of 20-30 spheres. As neither (I) nor (II) happens one has to mention that there are other reasons beyond the density argument: Reason 1 inhibits the growth of small nuclei, and Reason 2 leads to growth beyond 30 spheres. How reason 1 and 2 depend on each other is here not important. The current version refers only to (I) and Reason 1. In the frame of CNT (surface vs volume) this was enough, but now CNT is not present.} \MS{\it I think as a motivation for figure 4b and c, I is sufficient. Discussing II goes more towards arguing about CNT again.}

We thank \foreignlanguage{ngerman} {Markus Benderoth, Thomas Eggers, Tilo Finger, Kristian Hantke, Wolf Keiderling, Udo Krafft}, and Vishnu Natchu for technical support and discussions. The project was financed by grants from German Academic Exchange Service (DAAD), German Research Foundation (DFG) STA 425/24, Cluster of Excellence Engineering of Advanced Materials, and in part by NSF DMS-1208941, NSF DMS-1509088 and Robert A. Welch Foundation Grant F-0805.

\clearpage
\newpage
\onecolumngrid
\appendix*
\setcounter{figure}{0}  
\renewcommand{\thefigure}{S\arabic{figure}} 
\setcounter{page}{1}  
\renewcommand{\thepage}{S\arabic{page}} 

\section{\Large Supplemental Material: Nucleation in sheared granular matter\\
\vspace{0.5cm}
\large Frank Rietz, Charles Radin, Harry L. Swinney, and Matthias Schr\"oter}
\vspace{\fill}

\section{\large A. movie}
\noindent Link to movie: \url{https://youtu.be/_oV-WwtW4Xo}\\
\\The movie shows the nucleation of spheres in the shear cell. For visualization the spheres that are in a crystalline state are shown in a rotating side view. Color indicates the crystal symmetry.  At the top right the current state is specified by the global packing fraction, shear cycle, and the number of crystalline spheres. The end of the plateau in the packing fraction is marked by the appearance of the first growing nucleus, which is encircled in the rotating animation and depicted at a constant viewing perspective on the lower right. The wire frame indicates the inner part of the cell, while the green and brown frames are parallel to the shear walls. The bottom of the interrogation volume is cut non-orthogonally because of optical accessibility.

\section{\large B. Precision of sphere coordinates}

\begin{figure}[hb!]
\subfloat{\includegraphics[width=0.6\textwidth]{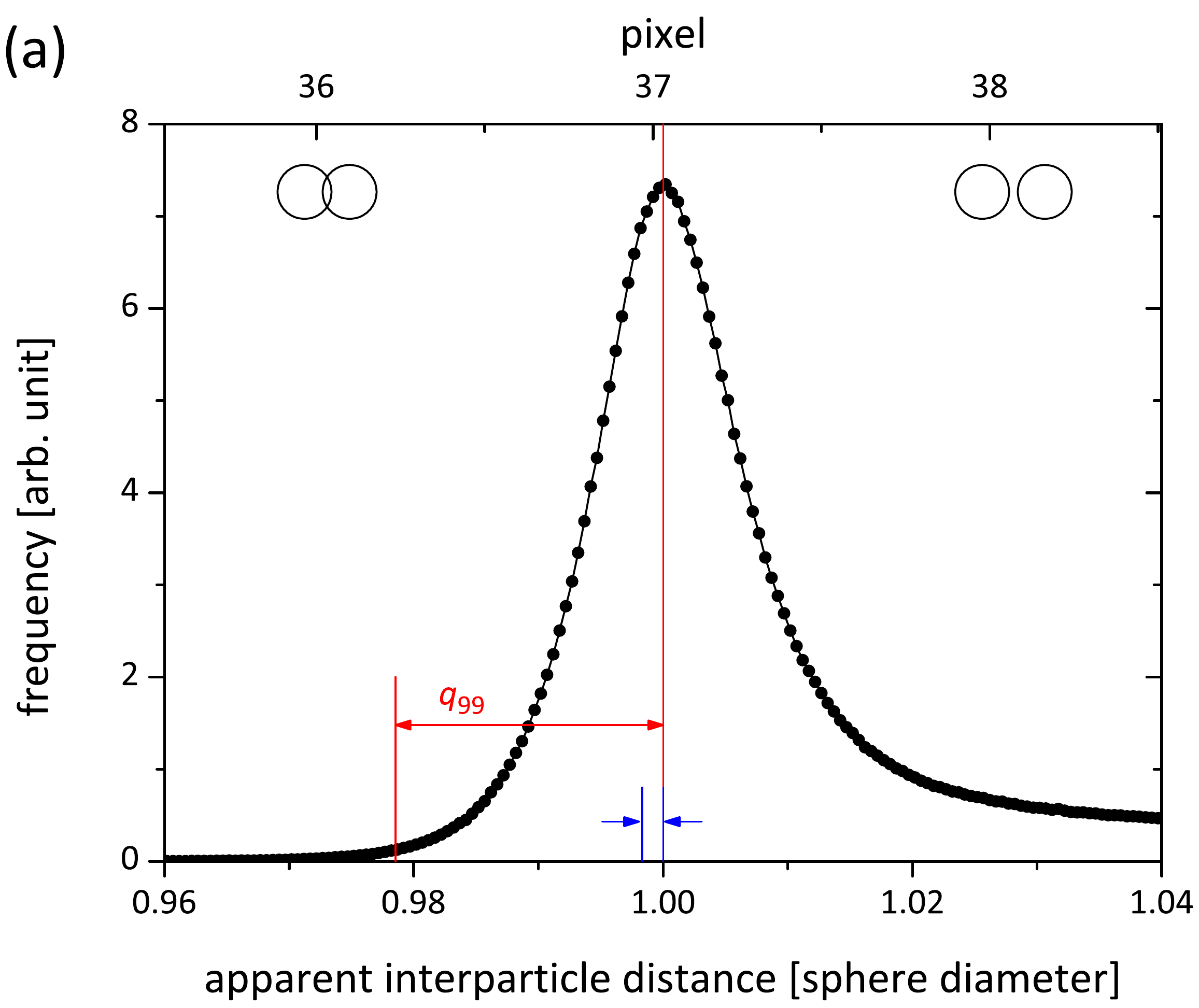}}
\hfill
\subfloat{\includegraphics[width=0.35\textwidth]{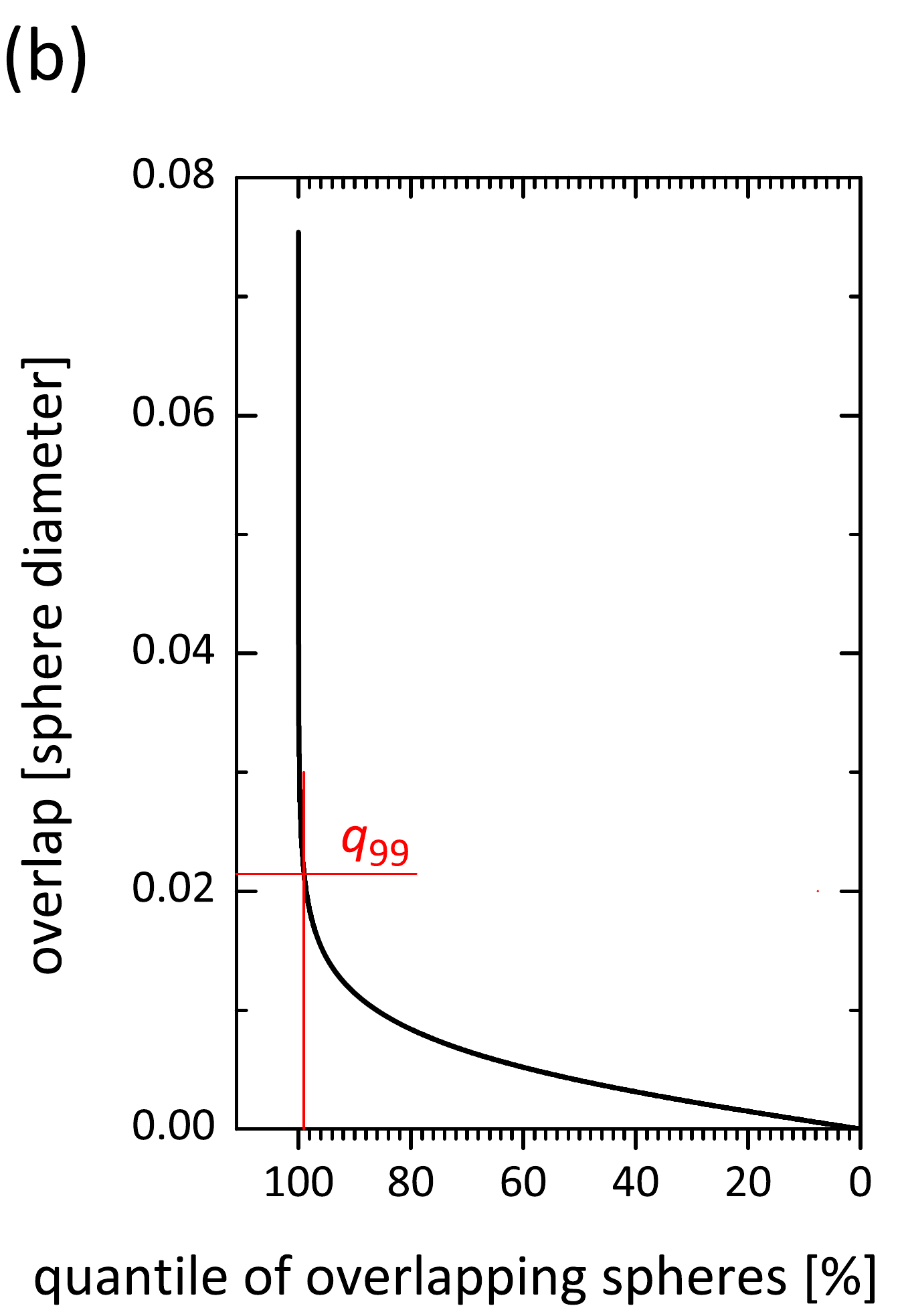}}
\caption{\label{fig_supp1}
(a)~The accuracy of the detected coordinates of the spheres with diameter $3\pm0.0025$\,mm is derived from the width of the first peak of the radial distribution function \cite{Weis2017_supp}.
For ideal monodisperse spheres and known coordinates there would be a step function at a distance of one particle diameter.
Under experimental conditions spheres can appear to overlap slightly due to imprecision in the particle positions.  The position of the peak maximum of the radial distribution function is used as the mean particle diameter and the width of the left shoulder characterizes the precision of the sphere coordinates.
(b)~The overlap length for a given percentile of overlapping contacts. Before counting the number of overlaps, spheres are sorted in increasing order, i.e., the smallest overlaps are first.
To exclude outliers the width of the left tail in (a) is defined by the largest of the 99\% smallest overlaps, the quantile $q_{99}$. The accuracy of the sphere positions is deduced to be 2\% of a diameter.  The blue arrows in (a) indicate overlap corresponding to the manufacturer's uncertainty in the diameter of the spheres. The coordinate error due to size tolerance is negligible compared to the inaccuracy of the detection method.  The evaluation includes $2.45 \times 10^6$ calculated sphere coordinates.
}\end{figure}

\clearpage
\section{\large C. Absence of icosahedral symmetry}

\begin{figure}[hb!]
\includegraphics[width=\columnwidth]{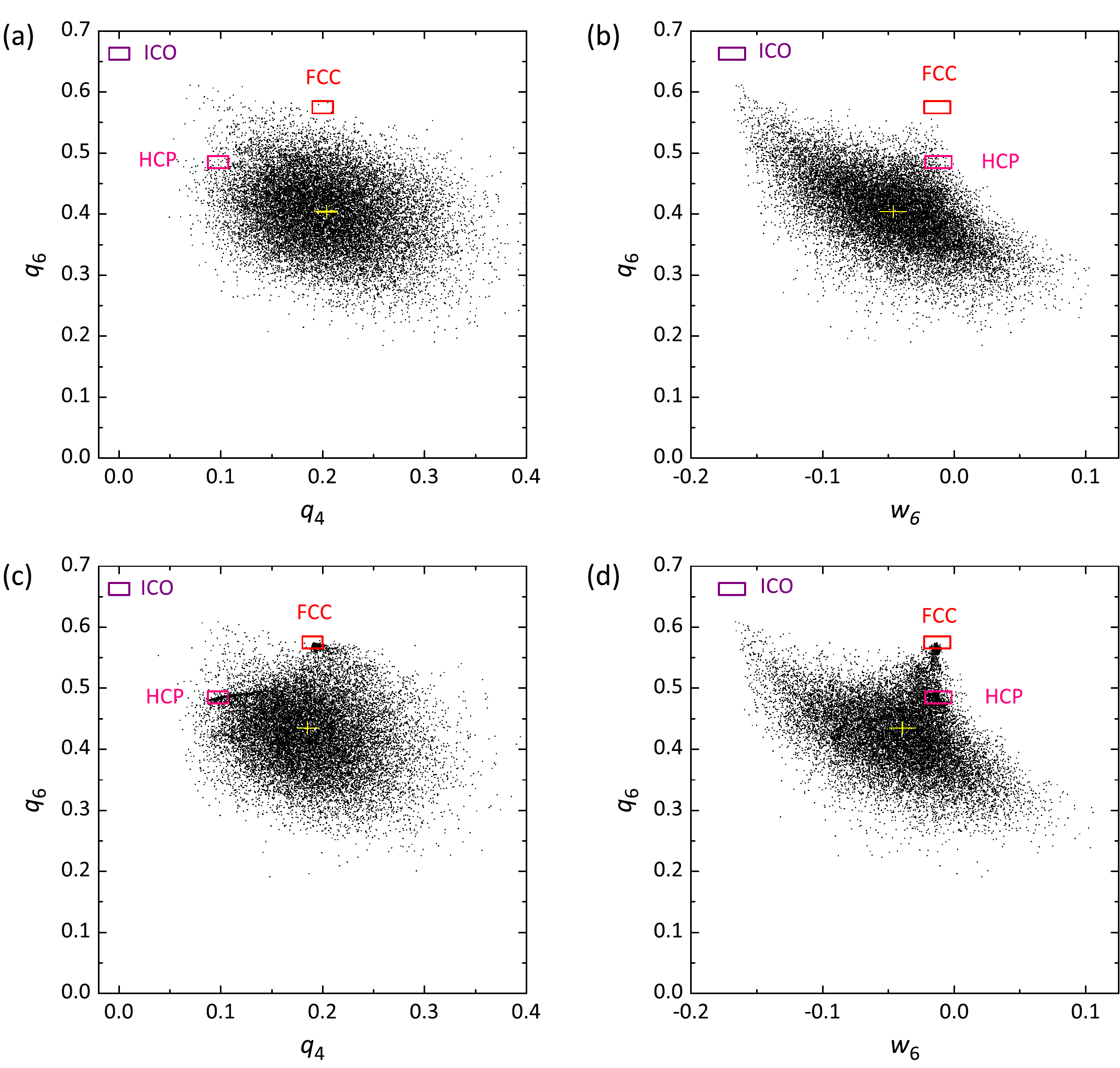}
\caption{\label{fig_supp12}
An absence of icosahedral symmetry is indicated by these two-dimensional mappings of the order parameters $q_6$ vs. $q_4$ and $q_6$ vs. $w_6$, where the order parameters $q_4$, $q_6$, and $w_6$ are calculated from the definition in \cite{Steinhardt1983_supp} together with the requirement that the Voronoi neighbors are weighted by the areas of their common Voronoi faces \cite{Mickel2013_supp} (see Fig.~\ref{fig_supp2}). The initial states are shown in (a) and (b), and the final states are shown in (c) and (d). Values for ideal structures of icosahedral (ICO), face-centered cubic (FCC) and hexagonal close packing (HCP) are located at the centers of the boxes, and the crosses indicate the mean over all spheres in the packing. Using the order parameter values together with a minimum Voronoi density $\phi_{local}$ (see Fig.~\ref{fig_supp2}), we conclude that there is no evidence for icosahedral symmetry, that is, there are no spheres possessing neighbors at the icosahedral positions. This contrasts with the icosahedral ordering that has been found for deeply cooled frictionless colloidal spheres (see Fig.~4 in \cite{Leocmach2012_supp}).
}\end{figure}

\clearpage
\section{\large D. No qualitative influence of different thresholds for crystal classification} 

\begin{figure}[hb!]
\includegraphics[width=0.7\columnwidth]{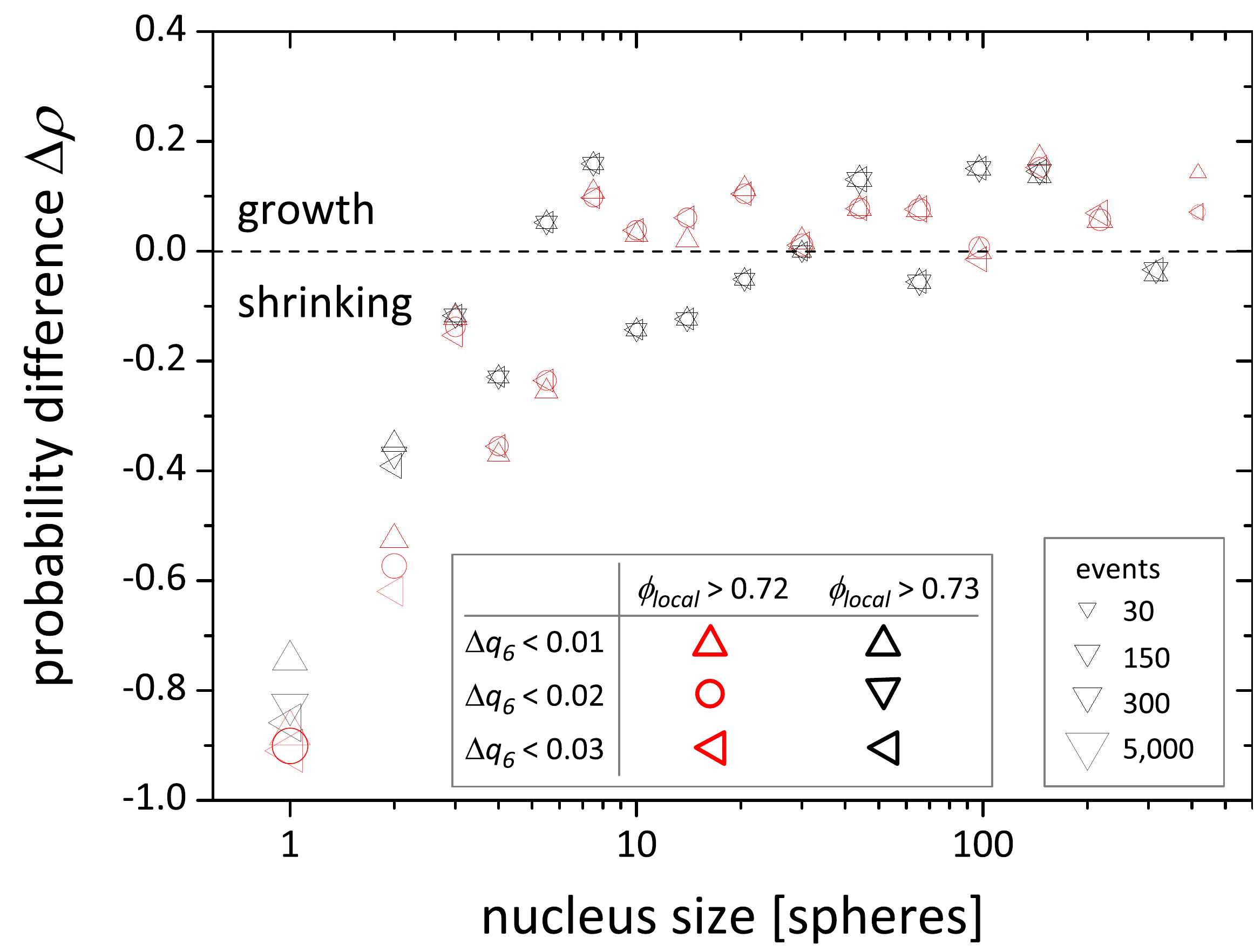}\caption{\label{fig_supp2}
Combinations of different thresholds for $\Delta q_6$ and $\phi_{local}$ do not change the general picture of preferred dissolution of nuclei with fewer than approximately ten spheres. We classify spheres as crystalline if $\phi_{local}>0.72$ and $\Delta q_6<0.02$, i.e., if $q_6$ is either in the range $q_6$(FCC)=0.575$\pm$0.02 or $q_6$(HCP)=0.485$\pm$0.02 (circle symbol). The rotationally invariant parameter $q_6$ is used in the literature to characterize the local order of a sphere by considering the relative positions of its surrounding particles \cite{Steinhardt1983_supp}. In $q_6$ neighbors are weighted by the area of the Voronoi faces that they share with the central sphere \cite{Mickel2013_supp}; thus close neighbors are more influential than distant ones. Symbol size indicates  number of observations.
}\end{figure}

\begin{figure}[hb!]
\includegraphics[width=\columnwidth]{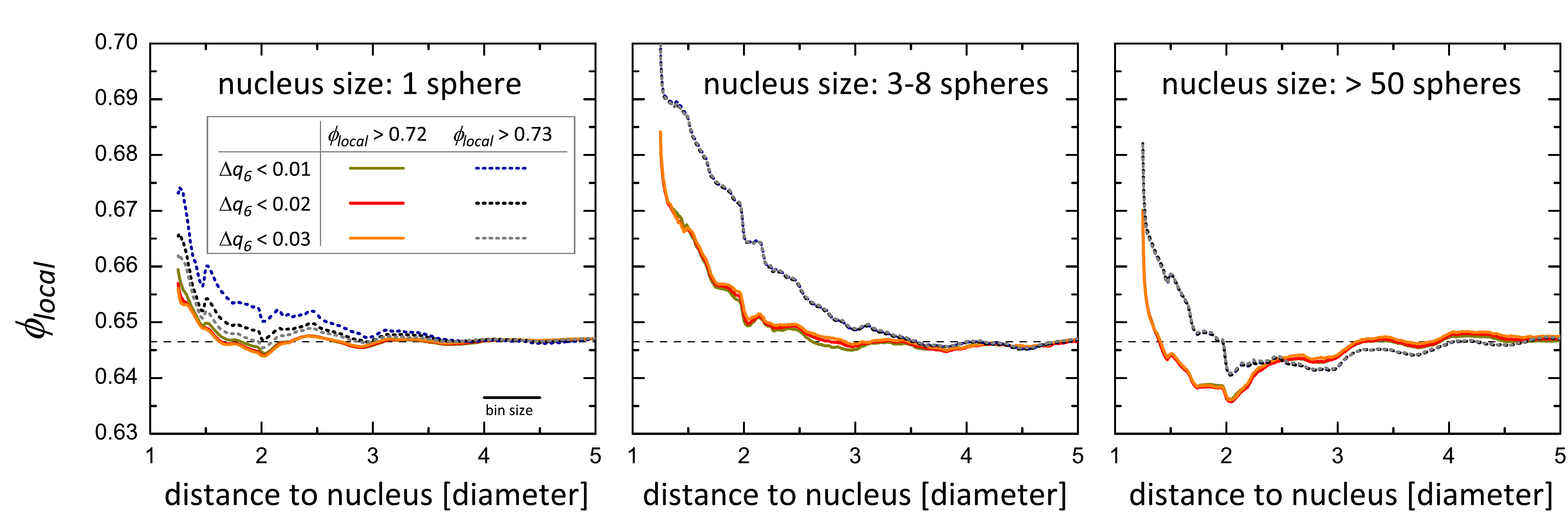}\caption{\label{fig_supp3}
Combinations of different thresholds do not change the general picture of an absence of zones of lower density around subcritical nuclei. The red solid line corresponds to the parameter value used in the paper.  To exclude possible interactions between different nuclei, only nuclei separated by at least 6 particle diameters have been included in the analyses here and in Fig.~\ref{fig4}(b) and Fig.~\ref{fig4}(c), and in Figs.~\ref{fig_supp5}, \ref{fig_supp9}, \ref{fig_supp11}, and \ref{fig_supp7}.
}\end{figure}

\clearpage
\section{\large E. No qualitative influence of a different crystal definition 1:\\Geometry of Voronoi cells}

\begin{figure}[hb!]
\includegraphics[width=0.7\columnwidth]{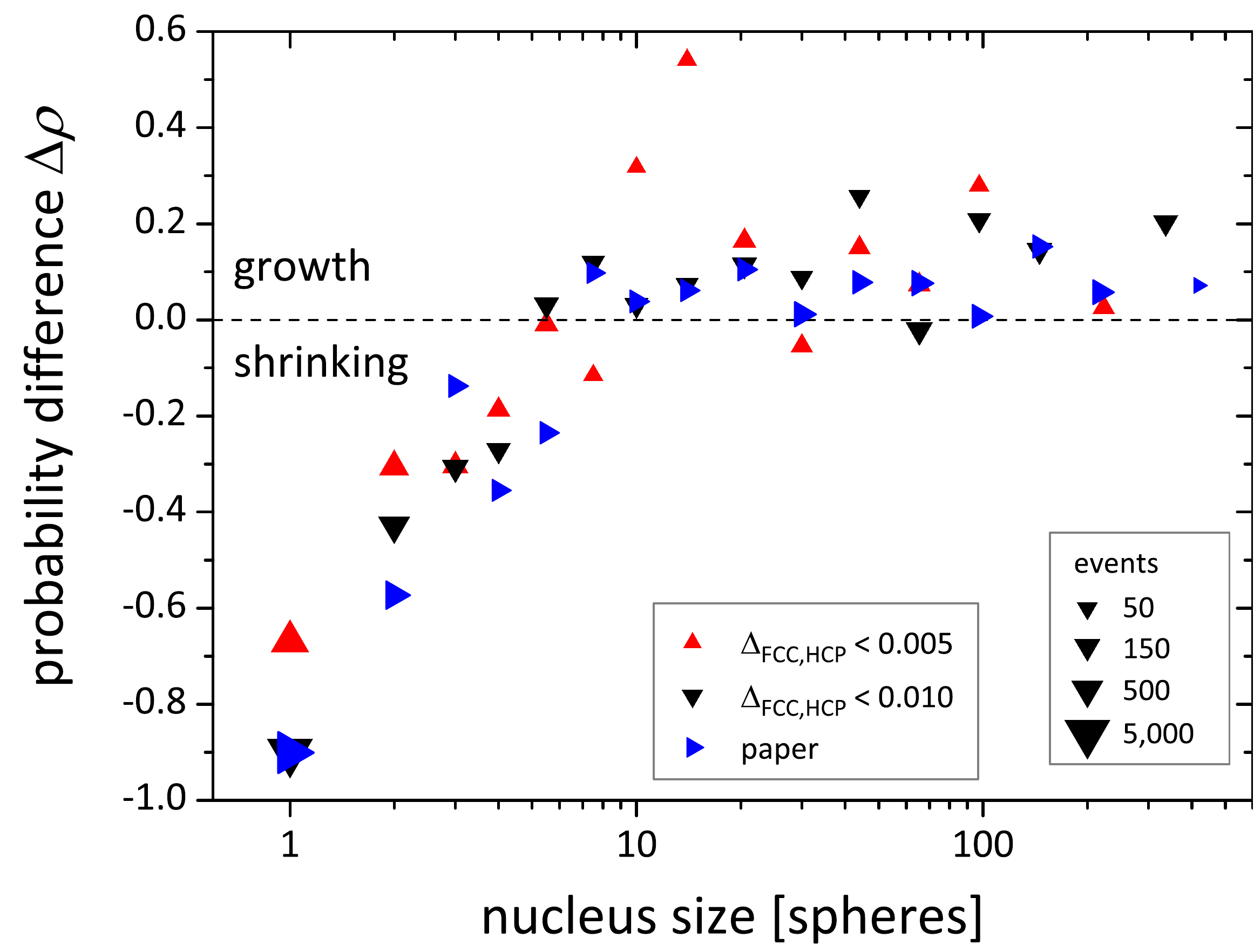}
\caption{\label{fig_supp4}
The results obtained for the growth and shrinking rates are insensitive to the definition used for the local crystallinity, whether it is a combination of $\Delta q_6$ and $\phi_{local}$ as in our paper, or it is a rotationally invariant fingerprint calculated from the face normals of a Voronoi cell \cite{Kapfer2012_supp}. A sphere has either FCC or HCP symmetry if the squared difference between the fingerprint of the Voronoi cell and and the fingerprints for ideal FCC or HCP Voronoi cells, $\Delta_{\text{FCC,HCP}}$, is below a threshold, as shown for the two values 0.005 and 0.010. The symbol sizes indicate the number of observations.
}\end{figure}

\begin{figure}[hb!]
\includegraphics[width=\columnwidth]{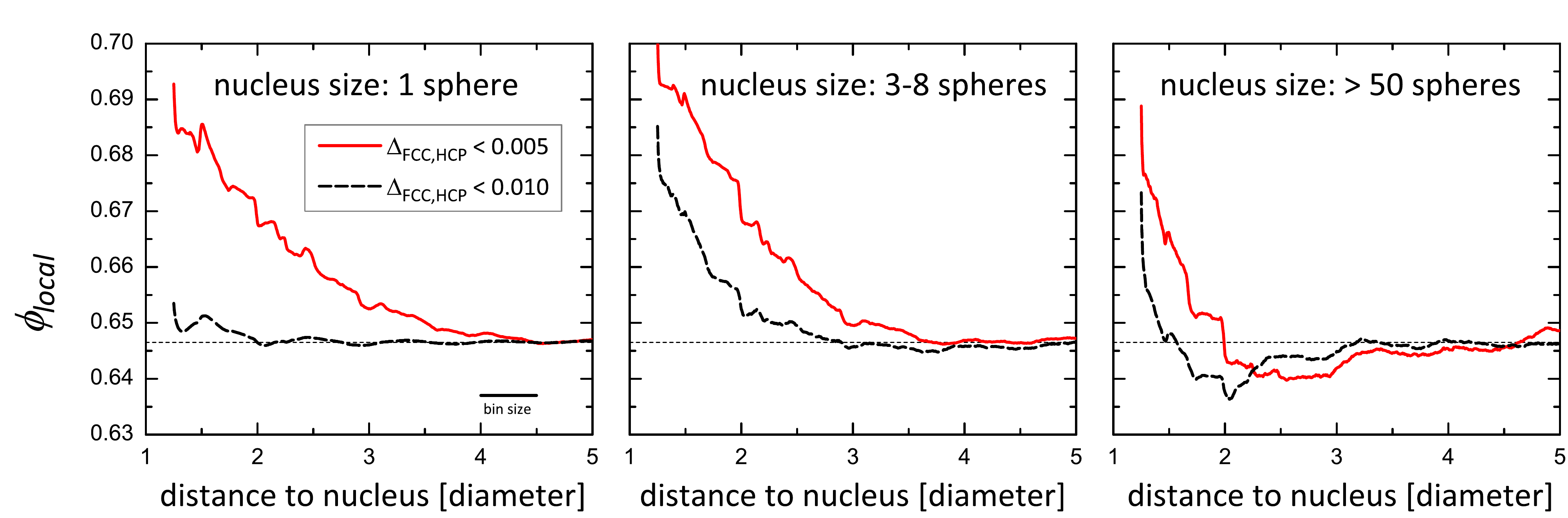}\caption{\label{fig_supp5}
Using $\Delta_{\text{FCC,HCP}}$ as a measure of local order, we find that the local volume fraction dependence on distance from a nucleus is qualitatively similar to that found in Fig.~\ref{fig_supp3}. Further, changing the threshold from 1.05 to 1.10 does not make a qualitative change in the properties at the interface.
}\end{figure}

\clearpage
\section{\large F. No qualitative influence of a different crystal definition 2:\\Common neighbor analysis}

\begin{figure}[hb!]
\includegraphics[width=0.7\columnwidth]{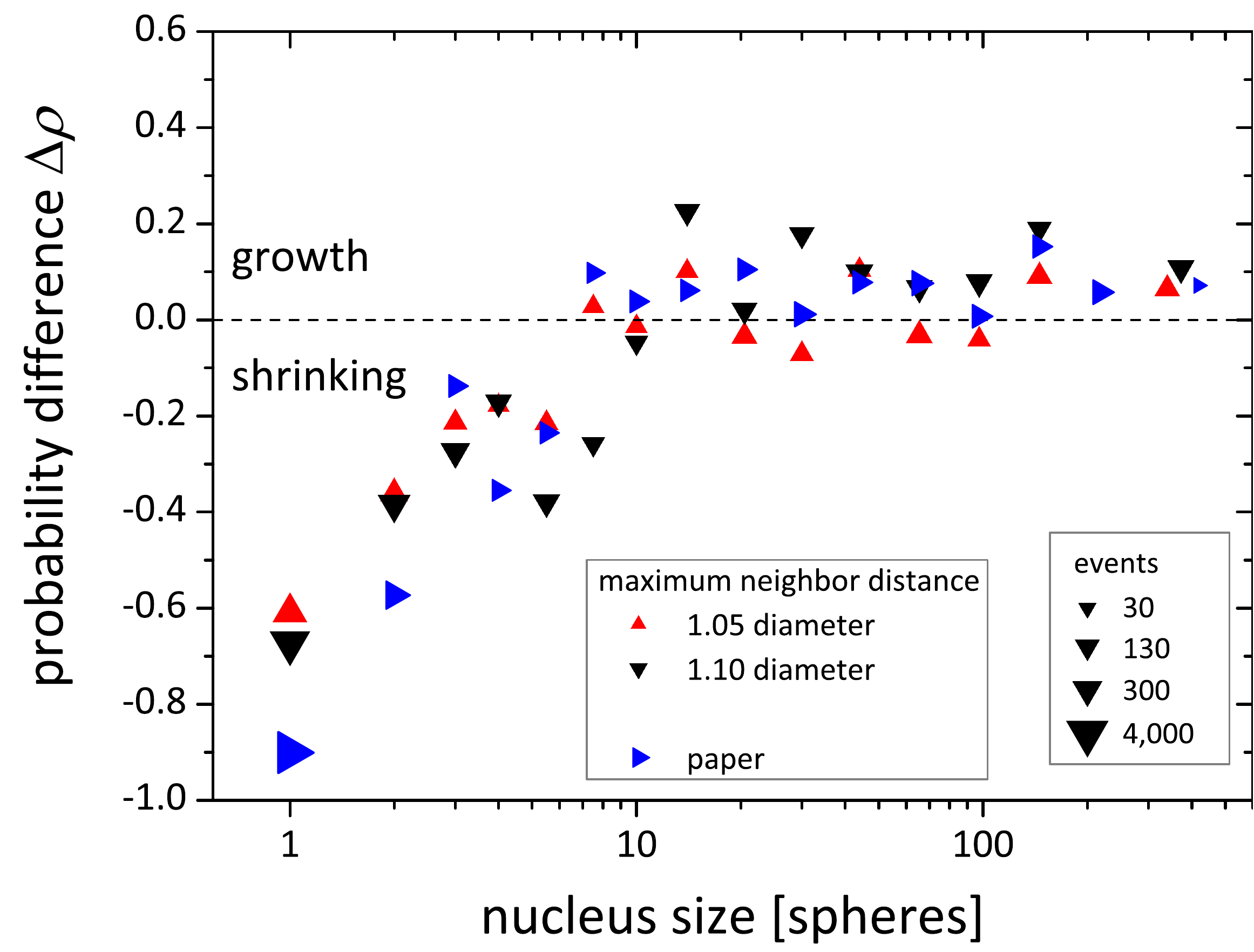}\caption{\label{fig_supp8}
The qualitative picture of the growth rates is unchanged if local crystallinity is detected by a common neighbor analysis  \cite{Faken1994_supp}, which is a widely used tool for structure identification. Among the neighbors a graph is constructed where mutual spheres are checked to determine if they are closer than a maximum distance. For this critical value the intermediate distance of the first and second shells of FCC, HCP, about 1.2~diameter, was suggested \cite{Stukowski2012_supp}. Using the method with this value results in abundant false positive detections; therefore, we use smaller thresholds of 1.05 and 1.10~diameter. A sphere is classified crystalline if the discrete signature derived from the graph is in accord with the signatures of either FCC or HCP. In the version of common neighbor analysis used, spheres with a partly crystalline neighborhood, such as on the surface of a crystallite, are classified as amorphous.
}\end{figure}

\begin{figure}[hb!]
\includegraphics[width=\columnwidth]{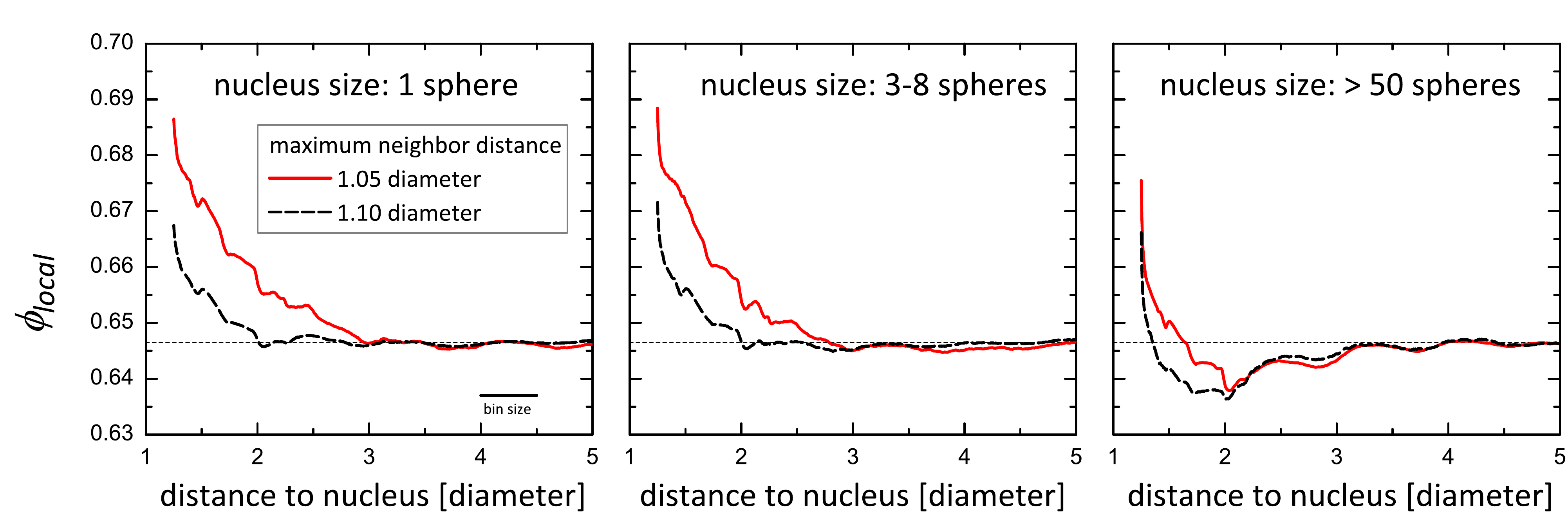}\caption{\label{fig_supp9}
Defining local order by a common neighbor analysis does not make a qualitative change of properties at the nucleus interface.
}\end{figure}

\clearpage

\section{\large G. No qualitative influence of a different crystal definition 3:\\Bond angle analysis}

\begin{figure}[hb!]
\includegraphics[width=0.7\columnwidth]{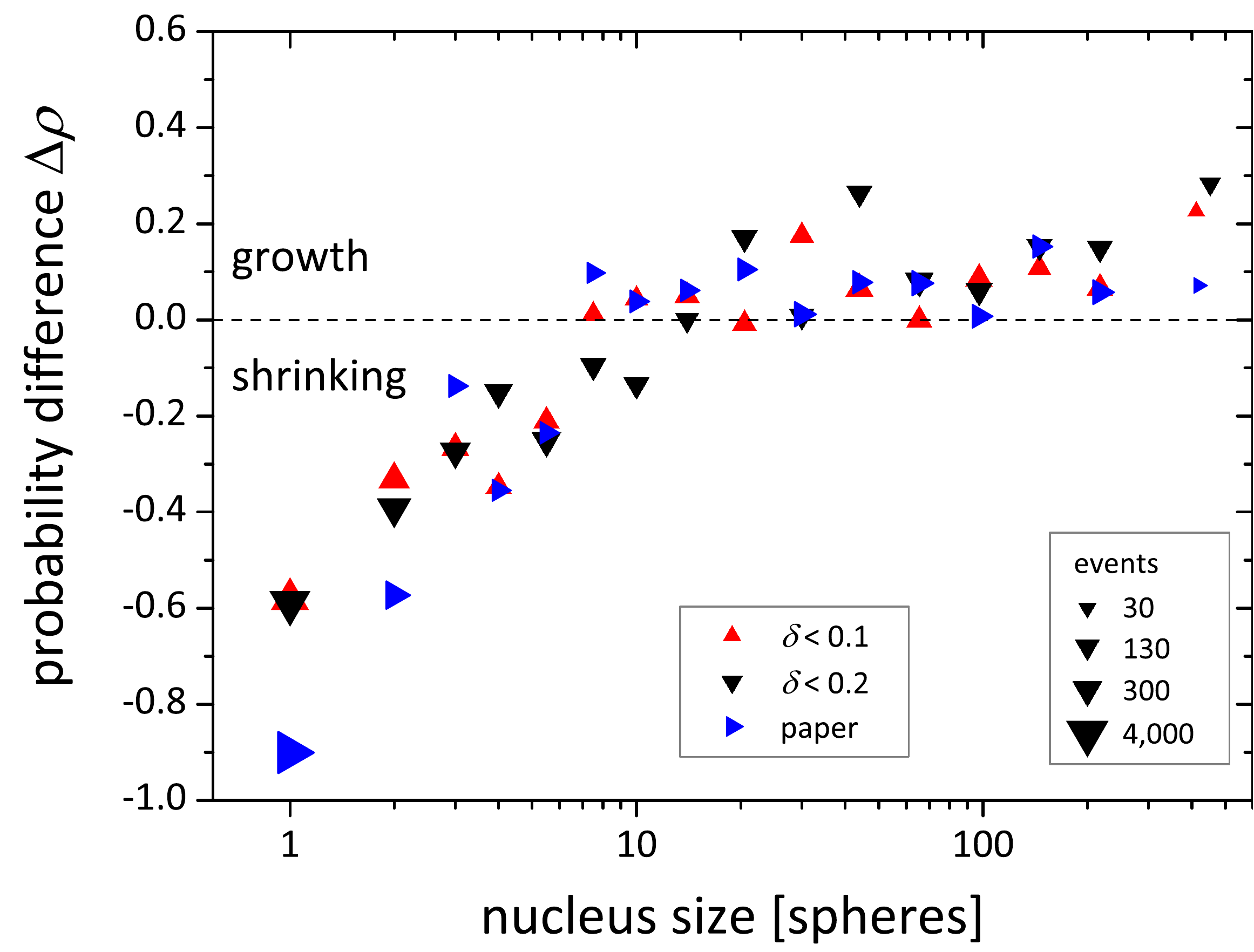}\caption{\label{fig_supp10}
Determination of local crystallinity by another method -- bond angle analysis -- does not qualitatively change the results for the growth and shrinking of nuclei. Here the angles formed by the central sphere with 2 of the 12 nearest neighbors are compared with the angles for ideal FCC and HCP. If the sum of $\binom{12}{2}=66$ squared angle differences (for angles in radians) are below the threshold $\delta$, the central sphere is classified as crystalline. Before comparison the angles must be sorted \cite{Bargiel2001_supp}.
}\end{figure}

\begin{figure}[hb!]
\includegraphics[width=\columnwidth]{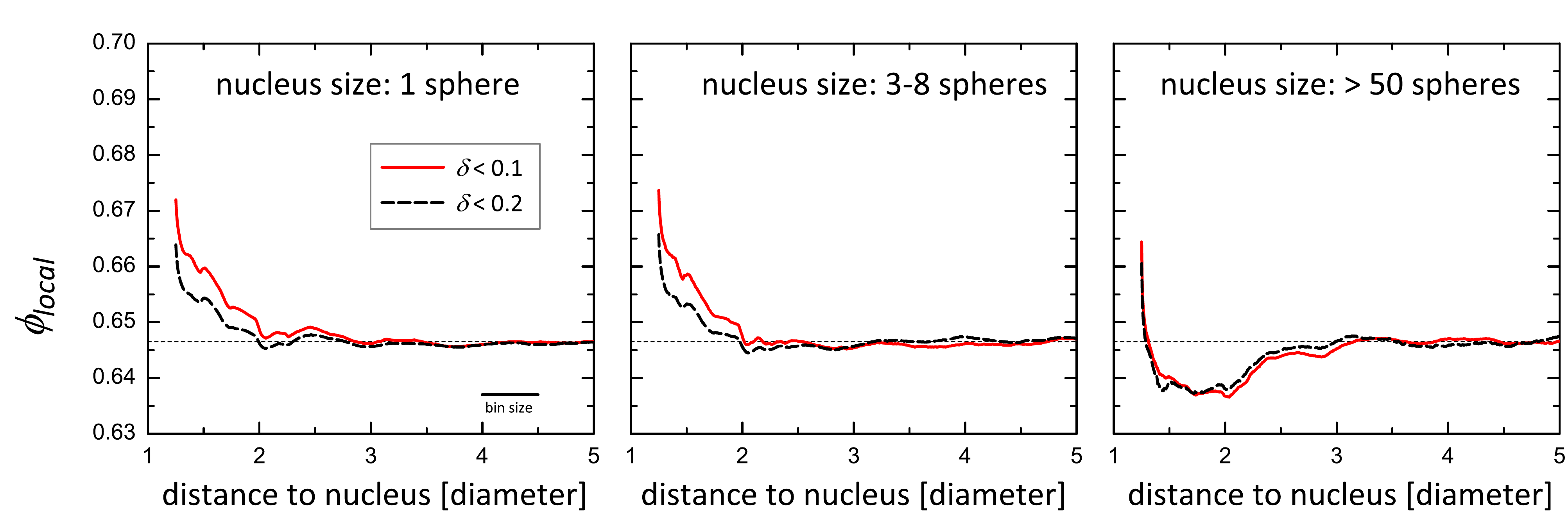}\caption{\label{fig_supp11}
Defining local order by a bond angle analysis does not yield a qualitative change of properties at the nucleus interface.
}\end{figure}

\clearpage
\section{\large H. Influence of bin size on interface behavior}

\begin{figure}[hb!]
\includegraphics[width=0.765\columnwidth]{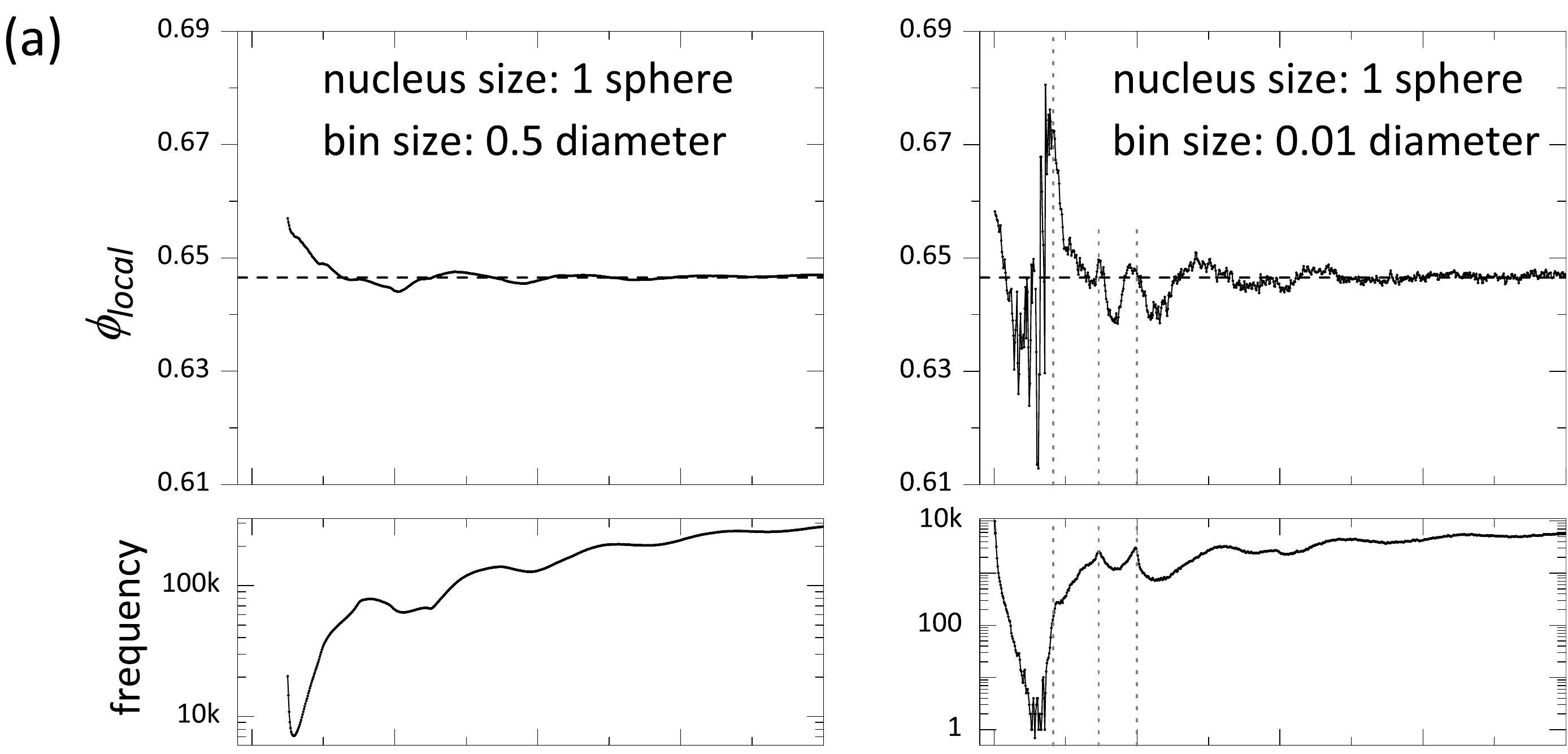}
\includegraphics[width=0.765\columnwidth]{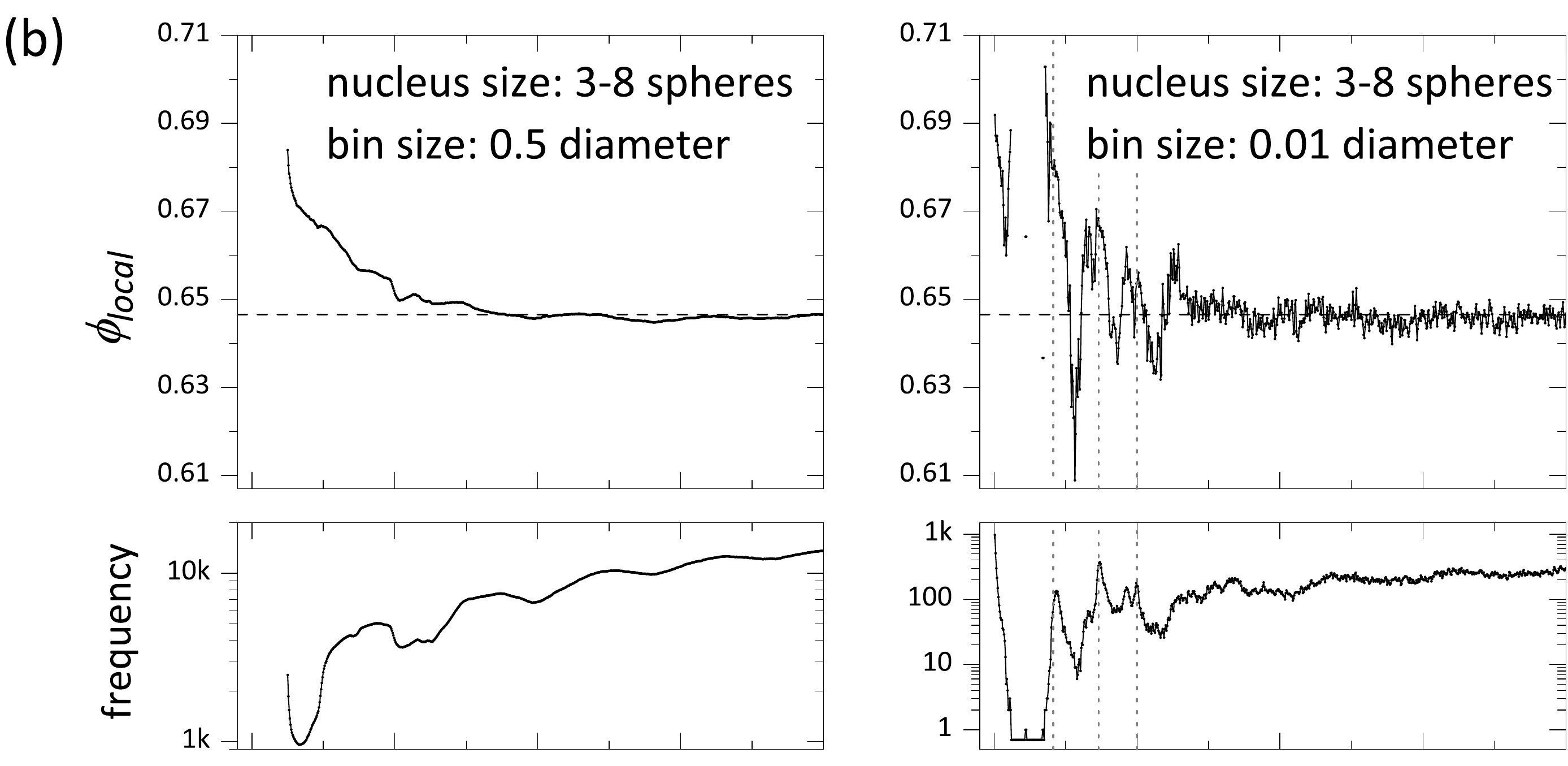}
\includegraphics[width=0.765\columnwidth]{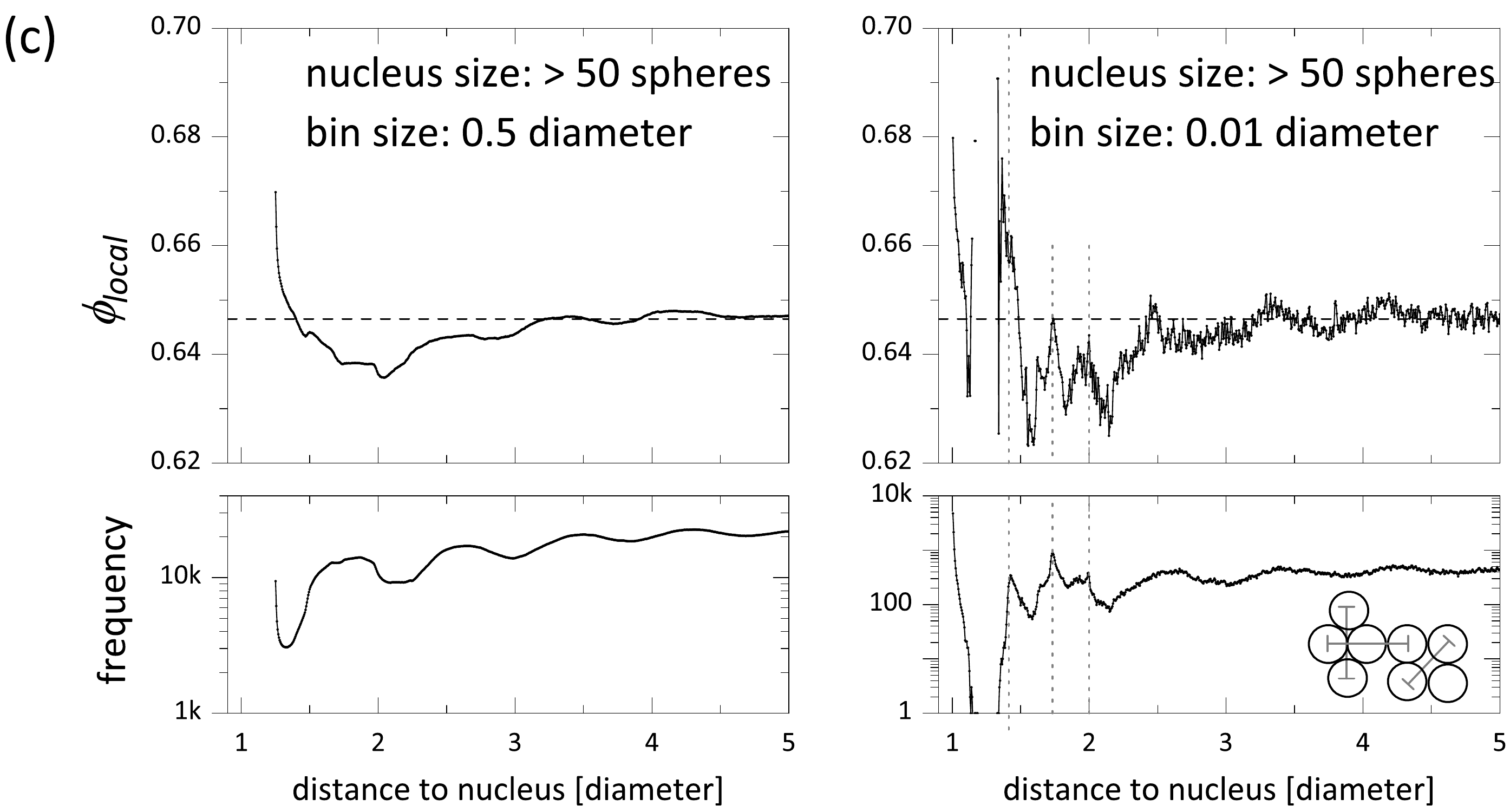}\caption{\label{fig_supp6}
Influence of bin size on $\phi_{local}$ at the nucleus interface. Left column: bin size 0.5~diameter, as used in the paper; right column: bin size 0.01~diameter. The bin step is always 0.005~diameter.  The pronounced peaks in the number densities at $\sqrt{2}$, $\sqrt{3}$, and 2~diameters correspond to the elementary planar distances illustrated on the lower right in (c).
}\end{figure}

\section{\large I. No qualitative difference in the comparison of growing and shrinking nuclei}

\begin{figure}[hb!]
\includegraphics[width=\columnwidth]{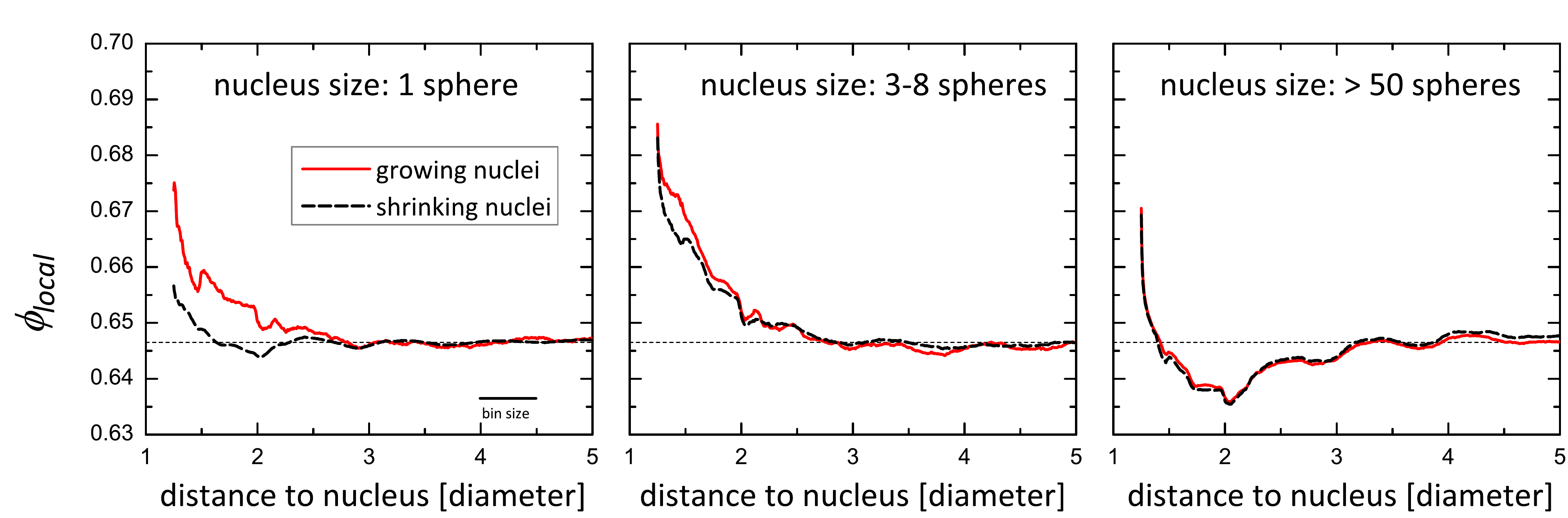}\caption{\label{fig_supp7}
The nuclei are split in groups of growing and shrinking nuclei. There is no qualitative difference in $\phi_{local}$ at the nucleus interface.
}\end{figure}

\clearpage\setlength{\parindent}{0pt}


\begin{thebibliography}{10}

\bibitem{Finney2013}%Bernal’s road to random packing and the structure of liquids
J. L. Finney, Philos. Mag. {\bf 93}, 3940 (2013).

\bibitem{Pusey1986}%Phase behaviour of concentrated suspensions of nearly hard colloidal spheres
P. N. Pusey and W. v. Megen, Nature (London) {\bf 320}, 340 (1986).

\bibitem{Gasser2001}%Real-space imaging of nucleation and growth in colloidal crystallization
U. Gasser, E. R. Weeks, A. Schofield, P. N. Pusey, and D. A. Weitz, Science {\bf 292}, 258 (2001).

\bibitem{Liber2013}%Dense colloidal fluids form denser amorphous sediments
S. R. Liber, S. Borohovich, A. V. Butenko, A. B. Schofield, and E. Sloutskin, PNAS {\bf 110}, 5769 (2013). 

\bibitem{Francois2013}%Geometrical Frustration in Amorphous and Partially Crystallized Packings of Spheres
N. Francois, M. Saadatfar, R. Cruikshank, and A. Sheppard, Phys. Rev. Lett. {\bf 111}, 148001 (2013).

\bibitem{Hanifpour2014_Hanifpour2015_Saadatfar2017}%Mechanical Characterization of Partially Crystallized Sphere Packings; Structural and mechanical features of the order-disorder transition in experimental hard-sphere packings;Pore configuration landscape of granular crystallization
M. Hanifpour, N. Francois, S. M. Vaez Allaei, T. Senden, and M. Saadatfar, Phys. Rev. Lett. {\bf 113}, 148001 (2014);  M. Hanifpour, N. Francois, V. Robins, A. Kingston, S. M. Vaez Allaei, and M. Saadatfar, Phys. Rev. E {\bf 91}, 062202 (2015); M. Saadatfar, H. Takeuchi, V. Robins, N. Francois, and Y. Hiraoka, Nat. Commun. {\bf 8}, 15082 (2017).

\bibitem{Bernal1964_Nicolas2000_Mueggenburg2005}%Growth of crystals from random closed packing; Compaction of a granular material under cyclic shear; Behavior of granular materials under cyclic shear
D. Bernal, K. R. Knight, and I. Cherry, Nature (London) {\bf 202}, 852 (1964); M. Nicolas, P. Duru, and O. Pouliquen, Eur. Phys. J. E {\bf 3}, 309 (2000); N. W. Mueggenburg, Phys. Rev. E {\bf 71}, 031301 (2005).

\bibitem{Panaitescu2012}%Nucleation and crystal growth in sheared granular sphere packings
A. Panaitescu, K. A. Reddy, and A. Kudrolli, Phys. Rev. Lett. {\bf 108}, 108001 (2012).

\bibitem{Tsai2003_Daniels2006}%Internal Granular Dynamics, Shear-Induced Crystallization, and Compaction Steps;Characterization of a freezing/melting transition in a vibrated and sheared granular medium
J.-C. Tsai, G. A. Voth, and J. P. Gollub, Phys. Rev. Lett. {\bf 91}, 064301 (2003); K. E. Daniels and R. P. Behringer, J. Stat. Mech., P07018 (2006).

\bibitem{Komatsu2015} Y. Komatsu and H. Tanaka, Phys. Rev. X {\bf 5}, 031025 (2015).

\bibitem{Chen2006_Slotterback2008}%Packing grains by thermal cycling;Correlation between Particle Motion and Voronoi-Cell-Shape Fluctuations during the Compaction of Granular Matter
K. Chen, J. Cole, C. Conger, J. Draskovic, M. Lohr, K. Klein, T. Scheidemantel, and P. Schiffer, Nature (London) {\bf 442}, 257 (2006); S. Slotterback, M. Toiya, L. Goff, J. F. Douglas, and W. Losert, Phys. Rev. Lett. {\bf 101}, 258001 (2008).


\bibitem{Knight1995_Richard2005}%Density relaxation in a vibrated granular material; Slow relaxation and compaction of granular systems
J. B. Knight, C. G. Fandrich, C. N. Lau, H. M. Jaeger, and S. R. Nagel, Phys. Rev. E {\bf 51}, 3957 (1995); P. Richard, M. Nicodemi, R. Delannay, P. Ribiere, and D. Bideau, Nat. Mater. {\bf 4}, 121 (2005).

\bibitem{Schroeter2005}%Stationary state volume fluctuations in a granular medium
M. Schr\"oter, D. I. Goldman,  and H. L. Swinney,  Phys. Rev. E {\bf 71}, 030301(R) (2005).

\bibitem{Scott1960_Scott1969_Finney1970}
G. D. Scott, Nature (London) {\bf 188}, 908 (1960); G. D. Scott and D. M. Kilgour, J. Phys. D {\bf 2}, 863 (1969); J. L. Finney, Proc. R. Soc. London, Ser. A {\bf 319}, 479 (1970). 

\bibitem{Berryman1983} J. G. Berryman, Phys. Rev. A {\bf 27}, 1053 (1983).

\bibitem{Kapfer2012}%Jammed spheres: Minkowski tensors reveal onset of local crystallinity
S. C. Kapfer, W. Mickel, K. Mecke, G. E. Schr\"oder-Turk, Phys. Rev. E {\bf 85}, 030301(R) (2012).

\bibitem{Anikeenko2007_Baranau2014}%Polytetrahedral Nature of the Dense Disordered Packings of Hard Spheres;Random-close packing limits for monodisperse and polydisperse hard spheres
A. V. Anikeenko and N. N. Medvedev, Phys. Rev. Lett. {\bf 98}, 235504 (2007); V. Baranau and U. Tallarek, Soft Matter {\bf 10}, 3826 (2014).



\bibitem{Royer2015}%Precisely cyclic sand: Self-organization of periodically sheared frictional grains
 J. R. Royer and P. M. Chaikin, PNAS {\bf 112}, 49 (2015).

\bibitem{Mickel2013}%Shortcomings of the bond orientational order parameters for the analysis of disordered particulate matter
W. Mickel, S. C. Kapfer, G. E. Schr\"oder-Turk, and K. Mecke, J. Chem. Phys. {\bf 138}, 044501 (2013).

\bibitem{Jin2010}%A first-order phase transition defines the random close packing of hard spheres
Y. Jin and H. A. Makse, Physica A {\bf 389}, 5362 (2010).

\bibitem{Wood1957_Alder1957_Hoover1968}%Preliminary Results from a Recalculation of the Monte Carlo Equation of State of Hard Spheres; Phase Transition for a Hard Sphere System; Melting Transition and Communal Entropy for Hard Spheres
W. W. Wood and J. D. Jacobson, J. Chem. Phys. {\bf 27}, 1207 (1957); B. J. Alder and T. E. Wainwright, J. Chem. Phys. {\bf 27}, 1208 (1957); W. G. Hoover and F. H. Ree, J. Chem. Phys. {\bf 49}, 3609 (1968).

\bibitem{Auer2004}%Numerical prediction of absolute crystallization rates in hard sphere colloids
S. Auer and D. Frenkel, J. Chem. Phys. {\bf 120}, 3015 (2004).

\bibitem{Sanz2011_Valeriani2012}%Crystallization Mechanism of Hard Sphere Glasses; From compact to fractal crystalline clusters in concentrated systems of monodisperse hard spheres; 
E. Sanz, C. Valeriani, E. Zaccarelli, W. C. K. Poon, P. N. Pusey, and M. E. Cates, Phys. Rev. Lett. {\bf 106}, 215701 (2011); C. Valeriani, E. Sanz, P. N. Pusey, W. C. K. Poon, M. E. Cates, and E. Zaccarelli, Soft Matter {\bf 8}, 4960 (2012).

\bibitem{Yanagishima2017}%Common mechanism of thermodynamic and mechanical origin for ageing and crystallization of glasses
T. Yanagishima, J. Russo, and H. Tanaka, Nat. Commun. {\bf 8}, 15954 (2017).

\bibitem{Sanz2014}%Avalanches mediate crystallization in a hard-sphere glass
E. Sanz, C. Valeriani, E. Zaccarelli, W. C. K. Poon, M. E. Cates, and P. N. Pusey, PNAS {\bf 111}, 75 (2014).

\bibitem{Hales2012}
T. C. Hales, {\it Dense Sphere Packings. A blueprint for formal proofs} (Cambridge University Press, Cambridge, 2012).

\bibitem{Radin2008_Aristoff2010}%Random Close Packing of Granular Matter;Random close packing in a granular model
C. Radin, J. Stat. Phys. {\bf 131}, 567 (2008); D. Aristoff and C. Radin, J. Math. Phys. {\bf 51}, 113302 (2010).

\bibitem{Dijksman2012}%Refractive index matched scanning of dense granular materials
J. A. Dijksman, F. Rietz, K. A. L\H{o}rincz, M. v. Hecke, and W. Losert, Rev. Sci. Instrum. {\bf 83}, 011301 (2012).

\bibitem{Raffel2007}
M. Raffel, C. E. Willert, S. T. Wereley, and J. Kompenhans, {\it Particle Image Velocimetry} (Springer, Berlin, 2007), p.160

\bibitem{supp} {For a movie see \url{https://youtu.be/_oV-WwtW4Xo} and for additional evaluations see pp.~S1-S8.}

\bibitem{Weis2017}%Analyzing X-Ray tomographies of granular packings
S. Weis and M. Schr\"oter, Rev. Sci. Instrum. {\bf 88}, 051809 (2017).


\bibitem{Faken1994_Bargiel2001_Stukowski2012_Tanaka2012_Leocmach2013}%Systematic analysis of local atomic structure combined with 3D computer graphics;Packing fraction and measures of disorder of ultradense irregular packings of equal spheres. II. Transition from dense random packing;Structure identification methods for atomistic simulations of crystalline materials
D. Faken and H. J\'onsson, Comput. Mater. Sci. {\bf 2}, 279 (1994); M. Bargie{\l} and E. M. Tory, Adv. Powder Technol. {\bf 12}, 533 (2001); A. Stukowski, Modell. Simul. Mater. Sci. Eng. {\bf 20}, 045021 (2012); H. Tanaka, Eur. Phys. J. E {\bf 35}, 113 (2012); M. Leocmach, J. Russo, H. Tanaka, J. Chem. Phys. {\bf 138}, 12A536 (2013).

\bibitem{Reinhart2017}% Machine learning for autonomous crystal structure identification
W. F. Reinhart, A. W. Long,  M. P. Howard,  A. L. Ferguson,  and  A. Z. Panagiotopoulos,  Soft Matter {\bf 13}, 4733 (2017).

\end{thebibliography}

\begin{thebibliography}{10}

\bibitem{Weis2017_supp}%Analyzing X-Ray tomographies of granular packings
S. Weis and M. Schr\"oter, Rev. Sci. Instrum. {\bf 88}, 051809 (2017).

\bibitem{Steinhardt1983_supp} % \bibitem{Steinhardt1983}%Bond-orientational order in liquids and glasses
P. J. Steinhardt, D. R. Nelson, and M. Ronchetti, Phys. Rev. B {\bf 28}, 784 (1983).

\bibitem{Mickel2013_supp}%Shortcomings of the bond orientational order parameters for the analysis of disordered particulate matter
W. Mickel, S. C. Kapfer, G. E. Schr\"oder-Turk, and K. Mecke, J. Chem. Phys. {\bf 138}, 044501 (2013).

\bibitem{Leocmach2012_supp} % Roles of icosahedral and crystal-like order in the hard spheres glass transition
M. Leocmach and H. Tanaka, Nat. Comm.  {\bf 3}, 974 (2012).

\bibitem{Kapfer2012_supp}%Jammed spheres: Minkowski tensors reveal onset of local crystallinity
S. C. Kapfer, W. Mickel, K. Mecke, G. E. Schr\"oder-Turk, Phys. Rev. E {\bf 85}, 030301(R) (2012).

\bibitem{Faken1994_supp}%Systematic analysis of local atomic structure combined with 3D computer graphics
D. Faken and H. J\'onsson, Comput. Mater. Sci. {\bf 2}, 279 (1994).


\bibitem{Stukowski2012_supp}%Structure identification methods for atomistic simulations of crystalline materials
A. Stukowski, Modell. Simul. Mater. Sci. Eng. {\bf 20}, 045021 (2012).

\bibitem{Bargiel2001_supp}%Packing fraction and measures of disorder of ultradense irregular packings of equal spheres. II. Transition from dense random packing
M. Bargie{\l} and E. M. Tory, Adv. Powder Technol. {\bf 12}, 533 (2001).
\end{thebibliography}
\end{document}